\documentclass[reprint,superscriptaddress,showpacs,aps,prl]{revtex4-1}

\usepackage{graphicx}
\usepackage{bm}
\usepackage{amsmath}

\begin{document}

\title{Non-Gaussian Parameter in ${\bm k}$-Dimensional Euclidean Space}

\author{Zihan Huang}
 \email{huangzih14@mails.tsinghua.edu.cn}
\affiliation{Department of Chemical Engineering, Tsinghua University, Beijing 100084, P. R. China}
\author{Gaoming Wang}
\affiliation{Department of Mathematical Sciences, Tsinghua University, Beijing 100084, P. R. China}
\author{Zhao Yu}
\affiliation{Department of Statistics, Columbia University, New York 10027, USA}

\date{\today}

\begin{abstract}
We generalize the non-Gaussian parameter, which is utilized to characterize the distinction of dynamics between realistic and Gaussian Brownian diffusions, in $k$-dimensional Euclidean space.
\end{abstract}


\maketitle

\section{I. INTRODUCTION}
It is commonly presumed that Fickian diffusion satisfies the Gaussian statistics. Namely, the stochastic displacements of the particles induced by the thermal jiggling follow a Gaussain (normal) distribution \cite{intro1}. However, in recent years, numerous experimental and simulation results indicate that non-Gaussian diffusion is more prevalent than expected, in which mean-square displacement (MSD) is still proportional to time (Fickian), such as the diffusion of colloids on phospholipid fluid tubules \cite{nongauss1}, and the heterogeneous dynamics of two-dimensional colloids in hexatic and solid phases \cite{nongauss2}. The non-Gaussian features of these Fickian diffusions call for a general perspective and corresponding physical descriptions.

A straightforward but effective way to quantify the heterogeneity of the non-Gaussian diffusion, is to examine the ratio defined as $\langle r^{2n}(t) \rangle / {\langle r^2(t) \rangle}^n$, where $r(t)$ represents the displacement at time $t$ and $\langle \cdot \rangle$ is the ensemble average. Through comparing the ratio of the realistic diffusion with that of the Gaussian one, the deviation from the Gaussian form that diffusion shows can be quantitatively identified. A non-Gaussian parameter $\alpha_{n,k}(t)$ is thereby defined for this purpose, where $k$ is the space dimensionality. In 3-dimensional (3D) space, this parameter is written as
\begin{equation}
    \alpha_{n,3}(t) = c_{n,3}\frac{\langle r^{2n}(t) \rangle}{{\langle r^2(t) \rangle}^n}-1
\end{equation}
where $c_{n,3}=3^n/(2n+1)!!$ is the non-Gaussian coefficient in 3D space \cite{3dpara}. A larger $|\alpha_n(t)|$ indicates a more heterogeneous dynamics of diffusion, and $\alpha_n(t)=0$ when diffusion follows the Gaussian statistics. However, the general form of this parameter (i.e., in $k$-dimensional Euclidean space) has not yet been obtained, which is the critical and urgent challenge we address in the present letter.

\section{II. ONE DIMENSION}
Firstly, it is essential to summarize the diffusion of Brownian particles in one dimension as the basis of the further generalization. We assume that the space is homogeneous, thereby the diffusivity $D$ remains constant. Afterwards, the self-part of the van Hove function [$G_s(x,t)$] \cite{vanhove} can be obtained by solving the one-dimensional diffusion equation as follow \cite{Fick},
\begin{equation}
    \frac{\partial G_s(x,t \mid x_0)}{\partial t}=D\frac{\partial^2G_s(x,t \mid
    x_0)}{\partial x^2},
\end{equation}
with the initial condition given by
\begin{equation}
    G_s(x,t=0 \mid x_0)=\delta(x-x_0),
\end{equation}
where $x(t)$ and $x_0$ are the positions of the particle at time $t$ and 0 respectively. $G_s(x,t)$ thereby follows the form
\begin{equation}
    G_s(x,t)=\frac{1}{\sqrt{4\pi Dt}}\exp\left(-\frac{(x-x_0)^2}{4Dt}\right).
\end{equation}

Considering the Galilean invariance, it is feasible to fix the initial position of the particle at the origin of the Cartesian coordinates. Therefore, $G_s(x,t)$ can be equivalently rewritten as
\begin{equation}
    G_s(r,t)=\frac{1}{\sqrt{4\pi Dt}}\exp\left(-\frac{r^2}{4Dt}\right),
\end{equation}
where $r(t)$ is the displacement of the particle as mentioned in section I. Hence, $G_s(x,t)$ represents the probability distribution function of the displacement. Then we have
\begin{align}
    {\langle r^2(t) \rangle}&={\int\limits_{-\infty}^{+\infty }{r^2G_s(r,t)dr}}=2Dt,\\
    {\langle r^{2n}(t) \rangle}&=\int\limits_{-\infty}^{+\infty}{r^{2n}G_s(r,t)dr}
    =(2n-1)!!(2Dt)^n.
\end{align}
Therefore, $c_{n,1}$ can be set as $1/(2n-1)!!$ in one dimensional space.

\section{III. GENERALIZATION IN ${\bm k}$-DIMENSION}
Under the assumption that the $k$-dimensional Euclidean space is homogeneous, the diffusions in all directions are independent and share the same diffusivity $D$. Let ${\bm r}=(r_1,r_2,\cdots,r_k)$ be the displacement vector. Then ${\bm r}$ follows a multivariate Gaussian distribution, and the joint van Hove function of ${\bm r}$ can be given by
\begin{equation}
    G_s({\bm r},t)=\frac{1}{(\sqrt{2\pi})^k\sqrt{|{\bm \Sigma}|}}{\rm exp}\left(-\frac{1}{2}{\bm r}^{\rm T}{\bm \Sigma}^{-1}{\bm r}\right),
\end{equation}
where ${\bm \Sigma}$ is the covariance matrix and expressed as
\begin{equation}
    {\bm \Sigma}=
    \left[
        \begin{array}{ccccc}
        2Dt & 0 & 0 & \cdots & 0\\
        0 & 2Dt & 0 & \cdots & 0\\
        0 & 0 & 2Dt & \cdots & 0\\
        \vdots & \vdots & \vdots & \ddots & \vdots\\
        0 & 0 & 0 & \cdots & 2Dt\\
        \end{array}
    \right]_{k \times k}.
\end{equation}
Then $G_s({\bm r},t)$ can be explicitly written as
\begin{align}
    G_s({\bm r},t)=\frac{1}{(\sqrt{4\pi Dt})^k}{\rm exp}\left(-\frac{\sum_{i=1}^k{r_i^2}}{4Dt}\right)
    =\prod\limits_{i=1}^k{G_s(r_i,t)}.
\end{align}
As a consequence, the MSD ${\langle r^2(t) \rangle}$ can be obtained by
\begin{align}
    {\langle r^2(t) \rangle}
    &={\int\limits_{-\infty}^{+\infty }{(r_1^2+r_2^2+\cdots+r_k^2)G_s({\bm r},t)dr_1dr_2\cdots dr_k}}\nonumber\\
    &=\sum\limits_{i=1}^k{\int\limits_{-\infty}^{+\infty }{r_i^2}\prod\limits_{j=1}^k{G_s(r_j,t)dr_j}}\nonumber\\
    &=\sum\limits_{i=1}^k\left({\int\limits_{-\infty}^{+\infty }{r_i^2}G_s(r_i,t)dr_i}\cdot\prod\limits_{j = 1\hfill\atop
    j \ne i\hfill}^k{\int\limits_{-\infty}^{+\infty}{G_s(r_j,t)dr_j}}\right)\nonumber\\
    &=\sum\limits_{i=1}^k{\int\limits_{-\infty}^{+\infty }{r_i^2}G_s(r_i,t)dr_i}=2kDt.
\end{align}
However, the derivation of ${\langle r^{2n}(t) \rangle}$ is more complex. Using the {\it multinomial theorem}, we have
\begin{align}
    &{\langle r^{2n}(t) \rangle}\nonumber \\
    &={\int\limits_{-\infty}^{+\infty}{(r_1^2+r_2^2+\cdots+r_k^2)^nG_s({\bm r},t)dr_1dr_2\cdots dr_k}}\nonumber\\
    &={\int\limits_{-\infty}^{+\infty }{
    \sum{\frac{n!}{l_1!l_2!\cdots l_k!}r_1^{2l_1}r_2^{2l_2}\cdots r_k^{2l_k}}
    G_s({\bm r},t)dr_1dr_2\cdots dr_k}}\nonumber \\
    &=\sum{\frac{n!}{l_1!l_2!\cdots l_k!}\int\limits_{-\infty}^{+\infty}{
    \prod\limits_{i = 1}^k{r_i^{2l_i}G_s(r_i,t)dr_i}}}\nonumber \\
    &=\sum{\frac{n!}{l_1!l_2!\cdots l_k!}\prod\limits_{i = 1\atop
    l_i \ne 0}^k{(2l_i-1)!!(2Dt)^{l_i}}},
\end{align}
where $l_1+l_2+\cdots+l_k=n$, and the sum is taken over all combinations of nonnegative integer indices $l_1$ through $l_k$ such that the sum of all $l_i$ is $n$. 

Eq. (12) can thereby be simplified as follow (See section {\it i} in Appendix for the proof),
\begin{align}
    {\langle r^{2n}(t) \rangle}&=(2Dt)^n\sum{n!\prod\limits_{i = 1\atop
    l_i \ne 0}^k{\frac{(2l_i-1)!!}{l_i!}}}\nonumber\\
    &=(2Dt)^n\frac{(2n+k-2)!!}{(k-2)!!}.
\end{align}
Thus, $c_{n,k}$ is defined as $c_{n,k}=(k-2)!!k^n/(2n+k-2)!!$, and the non-Gaussian parameter in $k$-dimensional Euclidean space is given by
\begin{equation}
    \alpha_{n,k}(t)=\frac{(k-2)!!k^n\langle r^{2n}(t) \rangle}{(2n+k-2)!!{\langle r^2(t) \rangle}^n}-1.
\end{equation}

\section{IV. Appendix}
\subsection{i. Proof for Eq. (13)}
We use the mathematical induction to prove that Eq. (13) is exact. The equation that should be examined can be written as
\begin{equation}
    \sum{\prod\limits_{i=1}^k{\frac{(2l_i-1)!!}{l_i!}}}=\frac{(2n+k-2)!!}{n!(k-2)!!}
\end{equation}
for $k,n=1,2,\cdots$, where the meaning of the sum is same with that in Eq. (12).\\
{\bf Basis}: Eq. (7) has demonstrated that Eq. (15) holds for $k=1$ and arbitrary $n$.\\
{\bf Inductive step}: We assume that Eq.(15) holds for $k\leq m$ and arbitrary $n$. When $k=m+1$, we have $l_1+l_2+\cdots+l_{m+1}=n$. The following relation can thereby be obtained by using the induction hypothesis when $l_{m+1}$ is fixed,
\begin{equation}
    \sum\limits_{l_{m+1} \atop {\rm is~ fixed}}{\prod\limits_{i=1}^m{\frac{(2l_i-1)!!}{l_i!}}}
    =\frac{(2n+m-2-2l_{m+1})!!}{(n-l_{m+1})!(m-2)!!},
\end{equation}
where $l_{m+1}=0,1,\cdots,n$. Meanwhile, we have
\begin{equation}
    \prod\limits_{i=1}^{m}{\frac{(2l_i-1)!!}{l_i!}}=
    \frac{l_{m+1}!}{(2l_{m+1}-1)!!}\prod\limits_{i=1}^{m+1}{\frac{(2l_i-1)!!}{l_i!}}.
\end{equation}
This leads to
\begin{equation}
    \sum\limits_{l_{m+1} \atop {\rm is~ fixed}}{\prod\limits_{i=1}^m{\frac{(2l_i-1)!!}{l_i!}}}
    =\frac{l_{m+1}!}{(2l_{m+1}-1)!!}\sum\limits_{l_{m+1} \atop {\rm is~ fixed}}{\prod\limits_{i=1}^{m+1}{\frac{(2l_i-1)!!}{l_i!}}}.
\end{equation}
Rearranging terms and using Eq. (16), we have
\begin{align}
    &\sum\limits_{l_{m+1} \atop {\rm is~ fixed}}{\prod\limits_{i=1}^{m+1}{\frac{(2l_i-1)!!}{l_i!}}}\nonumber \\
    &=\frac{(2l_{m+1}-1)!!}{l_{m+1}!}\cdot
    \frac{(2n+m-2-2l_{m+1})!!}{(n-l_{m+1})!(m-2)!!}.
\end{align}
Hence,
\begin{align}
    &\sum{\prod\limits_{i=1}^{m+1}{\frac{(2l_i-1)!!}{l_i!}}} =\sum\limits_{l_{m+1}=0}^n{\sum\limits_{l_{m+1} \atop {\rm is~ fixed}}{\prod\limits_{i=1}^{m+1}{\frac{(2l_i-1)!!}{l_i!}}}}\nonumber\\
    &=\sum\limits_{l_{m+1}=0}^n{\frac{(2l_{m+1}-1)!!}{l_{m+1}!}\cdot
    \frac{(2n+m-2-2l_{m+1})!!}{(n-l_{m+1})!(m-2)!!}}\nonumber \\
    &=\frac{1}{n!(m-1)!!}\times \frac{(m-1)!!}{(m-2)!!} \times \nonumber\\
    &\sum\limits_{l_{m+1}=0}^n{{n \choose l_{m+1}}(2l_{m+1}-1)!!(2n+m-2-2l_{m+1})!!},
\end{align}
where ${n \choose l_{m+1}}=n!/[l_{m+1}!\cdot(n-l_{m+1})!]$. 

Now the goal is to prove
\begin{align}
    &\sum\limits_{i=0}^n{{n \choose k}(2i-1)!!(2n+m-2-2i)!!}\nonumber \\
    &=(2n+m-1)!!\times\frac{(m-2)!!}{(m-1)!!}.
\end{align}
Based on Eq. (21), it is easy to find that Eq. (15) also holds for $k=m+1$. See section {\it ii} in Appendix for the proof of Eq. (21).

\subsection{ii. Proof for Eq. (21)}
First we have
\begin{equation}
    (1+x)^{-\frac{1}{2}}(1+x)^{-\frac{m}{2}}=(1+x)^{-\frac{1+m}{2}}.
\end{equation}
Note that
\begin{equation}
    (1+x)^\alpha=\sum_{k=0}^{\infty}{{\alpha \choose k}x^k},
\end{equation}
where $|x|<1$. Thereby Eq. (22) can be transformed into
\begin{equation}
    \left[\sum_{i=0}^{\infty}{{-\frac{1}{2} \choose i}x^i}\right]
    \left[\sum_{j=0}^{\infty}{{-\frac{m}{2} \choose j}x^j}\right]
    =\left[\sum_{n=0}^{\infty}{{-\frac{1+m}{2} \choose n}x^n}\right].
\end{equation}
Since
\begin{align}
    &\left[\sum_{i=0}^{\infty}{{-\frac{1}{2} \choose i}x^i}\right]
    \left[\sum_{j=0}^{\infty}{{-\frac{m}{2} \choose j}x^j}\right]\nonumber \\
    &=\sum\limits_{n=0}^{\infty}{\left[\sum_{i=0}^{n}{{-\frac{1}{2} \choose i}{-\frac{m}{2} \choose n-i}}\right]x^n},
\end{align}
we obtain
\begin{equation}
    \sum_{i=0}^{n}{{-\frac{1}{2} \choose i}{-\frac{m}{2} \choose n-i}}
    ={-\frac{1+m}{2} \choose n}
\end{equation}
by comparing the right sides of Eqs. (24) and (25). Particularly,
\begin{align}
    {-\frac{1}{2} \choose i}{-\frac{m}{2} \choose n-i} 
    &= (-1)^n\frac{(2i-1)!!}{2^ii!}\cdot\frac{(2n+m-2-2i)!!}{2^{n-i}(n-i)!(m-2)!!},\\
    {-\frac{1+m}{2} \choose n}&=(-1)^n\frac{(2n+m-1)!!}{2^nn!(m-1)!!}.
\end{align}
Then we have
\begin{equation}
    \sum\limits_{i=0}^{n}{\frac{(2i-1)!!(2n+m-2-2i)!!}{i!(n-i)!(m-2)!!}}
    =\frac{(2n+m-1)!!}{n!(m-1)!!}.
\end{equation}
Thus,
\begin{align}
    &\sum\limits_{i=0}^{n}{\frac{n!}{i!(n-i)!}(2i-1)!!(2n+m-2-2i)!!}\nonumber \\
    &=(2n+m-1)!!\times\frac{(m-2)!!}{(m-1)!!},
\end{align}
which proves Eq. (21).

\section{acknowledgments}
\begin{acknowledgments}
We thank Dr. Pengyu Chen and Dr. Guoxi Xu for helpful discussions.
\end{acknowledgments}

\bibliography{Bibliography}

\end{document}